\documentclass[a4paper,11pt]{article}
\usepackage{pos}

\title{Hyperloop – The ALICE analysis train system for Run 3}

\author*[a]{Raquel Quishpe}
\author[b]{Jan Fiete Grosse-Oetringhaus}
\author[b]{Raluca Cruceru}
\author[b]{Costin Grigoras}

\affiliation[a]{University of Houston, Department of Physics, \\
  3507 Cullen Blvd, Houston, Texas 77204, United States}

\affiliation[b]{CERN,\\
Espl. des Particules 1, 1211 Meyrin, Switzerland}

\emailAdd{raquel.quishpe@cern.ch}

\abstract{ALICE analyses mostly deal with large datasets using the distributed Grid infrastructure. In LHC running periods  1 and 2, ALICE developed a system of analysis trains (so-called “LEGO trains”) that allowed the user to configure analysis tasks (called wagons) that run on the same data. The LEGO train system builds upon existing tools: the ALICE analysis framework as well as the Grid submission and monitoring infrastructure. This centralized system improved the resource utilization and provided a graphical user interface (UI), in addition to bookkeeping functionalities. Currently, 90$\%$ of ALICE analyses use the train system. The ongoing major upgrade for LHC Run 3 will enable the experiment to cope with an increase of Pb-Pb collision data of two orders of magnitude compared to the Run 1 and 2 data-taking periods. In order to process this unprecedented data sample, a new computing model has been implemented, the Online-Offline Computing System (O$^{2}$). Analysis trains will also be the main workhorse for analysis in Run 3: a new infrastructure, Hyperloop, is being developed based on the successful concept of the LEGO trains. The Hyperloop train system includes a different and improved UI using modern responsive web tools, bookkeeping, instantaneous automatic testing, and the production of derived skimmed data. So far, about 800 Hyperloop trains have been successfully submitted to the Grid and ALICE analysis facilities using converted Run 2 data. An overview of the ALICE train system concept is given, highlighting the improvements of the new Hyperloop framework for analysis in Run 3.}

\FullConference{%
 
 The Ninth Annual Conference on Large Hadron Collider Physics - LHCP2021\\
 7-12 June 2021\\
 Online
}

\begin{document}
\maketitle

\section{Introduction}
The ALICE collaboration analyzes large datasets using the distributed Grid. For Run 1 and Run 2, the ALICE detector recorded around 10 PBs every year. In order to read the data from a single dataset, about $10^{5}$ files have to be read, containing data of the order of 50 TB. The creation and submission of analysis jobs to the Grid can be done with the ALICE analysis framework, and monitored with the ALICE monitoring service, MonALISA. If different users are running their analysis for the same dataset in the Grid by themselves, this implies loading  $10^{5}$ files for each user, which decreases the CPU efficiency and results in computing resources waste. Hence, an organized analysis framework was created, the LEGO (Lightweight Environment for Grid Operators) train system. 

The LEGO train system, offers a web-based user interface (UI) where ALICE users can declare their existing analysis task in the form of a wagon, and enable it for a dataset without dealing with the complexity of the Grid \cite{Zimmermann:2015owa}. Wagons that are enabled for the same dataset, are put together in a train run by a train operator, which then is tested and, if successful, submitted to the Grid. At the end of a train run, the results are merged, and users do not need to monitor the evolution of their analysis jobs submission. The LEGO train system has had a successful impact within the ALICE collaboration, currently, 90$\%$ of ALICE analyses use the train system, leading to the submission of 16000 trains and 172 million Grid jobs in 2020.

\section{From Run 2 to Run 3}
During the Long Shutdown 2 (LS2), the ALICE detector went through major upgrades allowing it to collect in Run 3 roughly 2 orders of magnitude more data than in Run 2. This led to the implementation of a new computing model, the Online-Offline Computing System  (O$^{2}$)\cite{Buncic:2015ari}, which is not compatible with the existing LEGO train system. Given the new analysis framework, as well as other needed upgrades, a new organized analysis framework has been developed for Run 3, the Hyperloop train system. In the long term, the LEGO trains will be phased out and replaced by the Hyperloop trains. To allow users to utilize Run 2 data with the new software O$^{2}$, the Run 2 data is being converted to the Run 3 data format, and can be handled with the Hyperloop train system. Hyperloop trains are submitted to the ALICE Grid as well as to ALICE Analysis Facilities where specific data for analysis is staged.

\section{Hyperloop train system design}
Following the LEGO trains solid concept, the Hyperloop framework is a tool to run and manage analysis trains on the ALICE Grid. It builds on the O$^{2}$ analysis framework, MonALISA and its Grid submission tools (LPM). The database and backend have been implemented within MonALISA plus a java-based model that allows bookkeeping of wagons and datasets, and stored in a PostgreSQL database. The GUI has been developed with a modern user-friendly library, React.js\cite{reactjs}.\\

In the LEGO train system, trains are organized by physics working group (PWG) (e.g. heavy flavour, correlations and flow, etc) and by dataset type (raw data and Monte Carlo) and collision system (p-p, p-Pb, Pb-Pb, Xe-Xe), resulting in about 87 active trains. All this information is contained in static web pages, which eventually got too large to be loaded and displayed efficiently. Furthermore, the large number of active trains has led to some train runs using the same dataset not being run together as they belonged to a different PWG trains.\\

The Hyperloop UI offers different personalized web pages for users and train operators, making the page load faster and easier to navigate. Trains are no longer divided per PWG, instead, the concept of an analysis is introduced. An analysis is created using a task manager, JIRA, where the analyzers who can edit the wagon configurations are defined. The usage of JIRA allows an easier management of analysis for approval in ALICE.

\begin{figure}[h!]
\centering
\includegraphics[width=.83\textwidth]{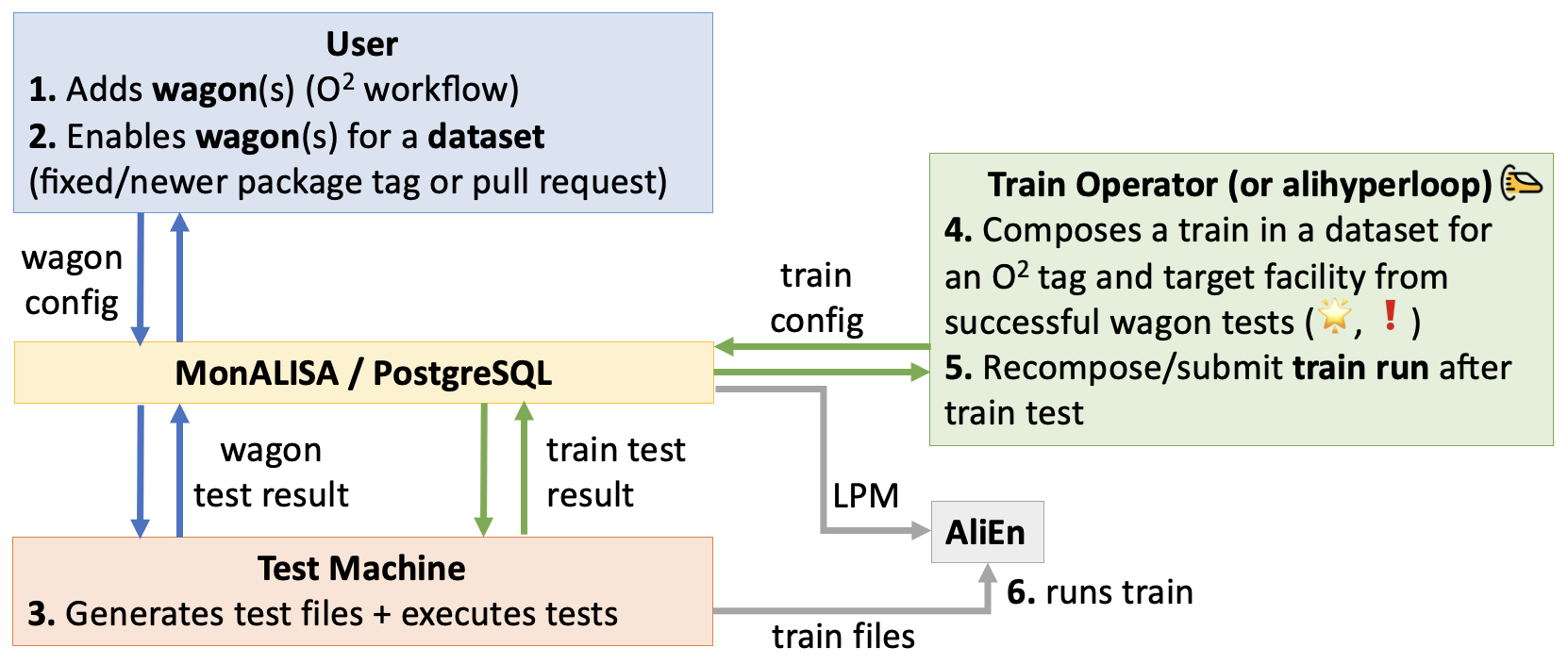}
\caption{Process to submit a Hyperloop train run.}
\label{process}
\end{figure}

Inside of an analysis, analyzers can create, configure and enable their wagons from an  O$^{2}$ workflow available in a daily release of the O$^{2}$ software (package tag). In addition to that, the analyzers also have the possibility to enable their wagon for the next release including changes contained in a GitHub pull request. Enabled wagons are automatically tested, and if enabled with a pull request, the wagon will be tested once the pull request has been included in a package tag, allowing the user to have an immediate feedback. If the wagon test is successful, a train operator can compose a train run for a dataset taking into account the required package tag, target facility maximum memory and wagons common dependencies. An algorithm has been developed returning the ideal wagon combination based on the individual wagon tests and the parameters mentioned before, hence, the train operator does not need to manually estimate the combination and can compose train runs more efficiently with optimal computing resource usage. Moreover, automatic train run composition can be scheduled for a dataset, so that the process previously mentioned can be automatically executed by the system. Figure \ref{process} summarizes the process of a hyperloop train submission. The fact that most of the actions are now automatic allows to have less train operators but who are experts within the framework.\\ 

\begin{figure}[h!]
\centering
\includegraphics[width=.92\textwidth]{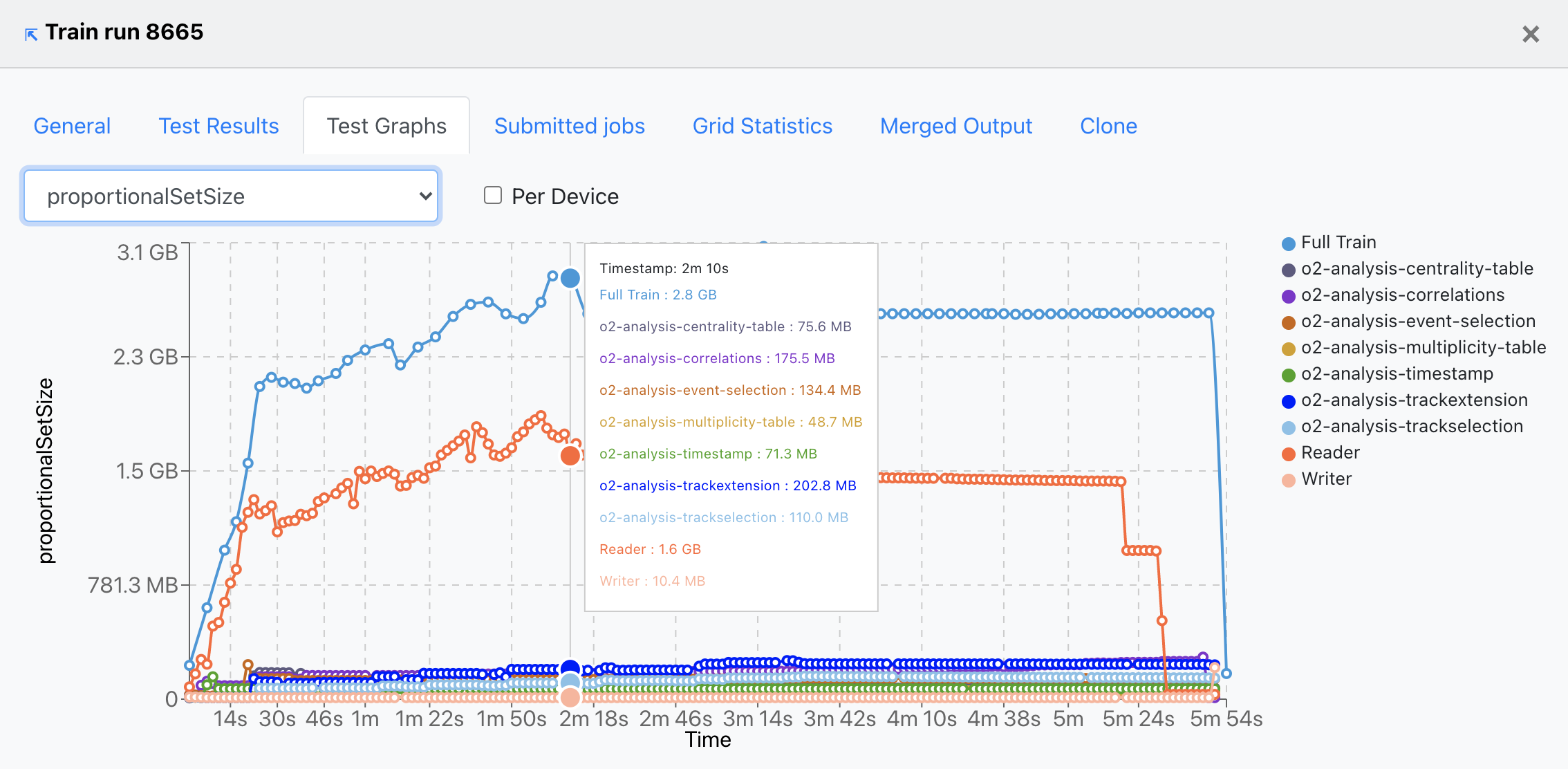}
\caption{Train run test performance result. The proportional set size memory of all workflow tasks in the train run are displayed.}
\label{result}
\end{figure}

Thanks to the bookkeeping feature, a change log web page has been developed for wagons and datasets, providing the configuration at a specific time. Wagon and train comparison pages have been created as well. Also, for wagons and train runs test results, interactive performance graphs and personalized notifications have been implemented (example in figure \ref{result}). 

\section{Current status of the Hyperloop train system}

A first version of the Hyperloop train system was available for the ALICE Analysis Challenge, which began on June 2020. So far, 24 analyses have been created for the Hyperloop framework, where 39 analyzers are actively using it. Since then, continuous improvements have been made to the Hyperloop framework following the improvements made to the O$^{2}$ software. Analyzers from different PWGs have been testing the new framework using converted Run 2 data. Their results have shown a 3-10 event throughput improvement from AliPhysics (LEGO trains) to O$^{2}$ (Hyperloop trains),  further optimization is foreseen. To date, more than 800 Hyperloop trains have been successfully submitted to the Grid and Analysis Facilities.

\section{Summary}

A modern and reactive organized trains framework, Hyperloop, is being successfully developed based on the solid concept and results from the LEGO train system. The Hyperloop train system has been designed to support the ALICE Run 3 software suite, O$^{2}$. Several improvements have been implemented, the most relevant being a unified train run submission, wagons and datasets bookkeeping, automatic wagon testing and automatic train composition and submission. Users from different PWGs have successfully tested the Hyperloop UI with converted Run 2 data.

\end{document}